\documentclass[10pt, conference, compsocconf]{IEEEtran}
\usepackage{graphicx,subfigure,wrapfig}
\usepackage{latexsym}
\usepackage{amsfonts}
\usepackage{amsmath}
\usepackage{amssymb}
\usepackage{epsfig}
\usepackage{stmaryrd}
\usepackage{xspace}
\usepackage{color}
\usepackage{B2}
\usepackage{graphicx}
\usepackage{epstopdf}
\DeclareGraphicsExtensions{.eps}
\IEEEoverridecommandlockouts

\usepackage{mdframed}
\DeclareFontFamily{OT1}{cmtt}{\hyphenchar \font=-1}
\DeclareFontFamily{\encodingdefault}{\ttdefault}{\hyphenchar\font=`\-}

\begin{document}
%
\title{Towards the Trustworthy Development of Active Medical Devices: A Hemodialysis Case Study\thanks{A. Mashkoor and M. Biro work at Software Competence Center Hagenberg, Hagenberg, Austria (e-mail: firstname.lastname@scch.at). The writing of this article is supported by the Austrian Ministry for Transport, Innovation and Technology, the Federal Ministry of Science, Research and Economy, and the Province of Upper Austria in the frame of the COMET center SCCH.}\thanks{The final publication is available at IEEE via https://doi.org/10.1109/LES.2015.2494459.}}
\author{\IEEEauthorblockN{Atif Mashkoor and Miklos Biro}}

\maketitle

\begin{abstract}
The use of embedded software is advancing in modern medical devices, so does its capabilities and complexity. This paradigm shift brings many challenges such as an increased rate of medical device failures due to software faults. In this letter, we present a rigorous ``correct by construction'' approach for the trustworthy development of hemodialysis machines, a sub-class of active medical devices. We show how informal requirements of hemodialysis machines are modeled and analyzed through a rigorous process and suggest a generalization to a larger class of active medical devices. 
\end{abstract}

\begin{IEEEkeywords}
Formal methods, requirements modeling, verification and validation, active medical devices, hemodialysis machines
\end{IEEEkeywords}

\section{Introduction}
\label{sec:intro}

An Active Medical Device (AMD) is a health-care device whose operation depends on a source of electrical energy or any source of power other than that directly generated by the human body or gravity and which acts by converting this energy \cite{ec93a}. Until recently, AMDs were mostly composed of mechanical components. However, recently embedded software has shown to have a determining impact on the consumer value of AMDs and their competitive differentiation. Consequently, according to the latest directive 2007/47/EC of the EU concerning medical devices \cite{ec07a}, a stand-alone software can also be considered as an AMD. The main reason of this change is that software lends itself to adaptation to individual requirements and requirements changes clearly much faster than hardware.

As AMDs become more and more software-intensive, due to the immaterial nature of software, their certification becomes a crucial issue. Certification is the process to determine the fitness of a device for public use. Despite the stringent mechanisms already in place to check the quality of medical products with respect to their safe operation, several incidents have been reported that were caused by software faults. Sandler et al. \cite{sandler10a} show that more than a fourth of the recalls of defective medical devices during the first half of 2010 were likely caused by software problems. Furthermore, because of the increased deployment of powerful programmable processors in medical devices, software-related recalls are on the rise \cite{wallace01a}.

Certification regimes have responded to these issues by proposing various medical software-related international standards and guidelines such as IEC 62304 \cite{iec06a} and FDA General Principles of Software Validation \cite{fda02a} or more general software-related standards such as IEC 61508-3 \cite{iec10a}. The standard IEC 62304 categories medical software into three classes. Class A software cause no injury or damage to health. Class B software cause no serious injury to health. Finally, class C software may cause serious injury or even death. Additionally, the standard also describes the software documents which are required to be produced for each class. Likewise, FDA also proposes several guidelines for medical software development. Some of the key documents of both IEC standards and FDA guidelines are: a requirements specification, a detailed architecture and design document, and an elaborated plan for early and rigorous verification and validation (V\&V) supporting all phases of software development life cycle. Although a device manufacturer has some flexibility in choosing V\&V principles, the manufacturer retains the ultimate responsibility for demonstrating that the software has been proven correct.

One of the key recommendations of these standards and guidelines is to adopt formal methods for the development of software-intensive critical systems. The use of formal methods is, in fact, ``highly recommended'' at higher Safety Integrity Levels (SILs). The safety integrity of a system can be defined as the probability of a safety-related system performing the required safety function under all of the stated conditions within a stated period of time. Highly recommended means that if the mentioned technique or measure is not used, then the rationale behind this choice has to be justified during safety planning and assessment. IEC 61508-3 further states that the confidence that can be placed in the software safety requirements specification, as a basis for safe software, depends on the rigor of techniques by which the desirable properties of the specification have been achieved. 

The main contribution of this letter is to show how a rigorous refinement-based approach can be applied to the trustworthy development of a sub-system of Hemodialysis (HD) machines, a sub-class of AMDs. The resulting formal model demonstrates an example of generalization, that how requirements can be rigorously specified and analyzed through a chain of refinements to be represented at various abstraction levels, to a larger class of AMDs. The approach also leads to a software safety requirements specification that guarantees correctness of the addressed aspects of behavior, supports verification of the specification based on systematic analysis, avoids intrinsic specification faults, reduces ambiguities in specification writing, and ultimately generates programming language code. The combined approach of requirements modeling, analysis and development based on techniques such as refinement, V\&V and translation, and tools such as proof checkers, model checkers, animation engines and code generators results in obtaining high-assurance and trustworthy AMDs. According to IEC 62304, HD machines are characterized as class C devices and the applied rigorous technique is particularly suitable for this class of AMDs as most of its components belong to a higher SIL.  

Formal methods have been used in the past for the development of various health-care devices such as cardiac-care products  \cite{chunxiao13a, mery13a, pajic12a} and infusion pumps \cite{bowen13a, masci13a}. However, the novelty of this work is that this is the first instance of the application of formal methods for the modeling, analysis and development of a sub-class of AMDs responsible for renal replacement therapy. We believe that our work will inspire the manufacturers of such systems to adopt formal paradigms for the safe and trustworthy development of variants of this domain.

\section{Methodology}
\label{sec:methodology}

The development of embedded software for AMDs is a complex process. The degree of complexity often leads to an artifact that, although requires a great amount of time, resources and attention to develop, yet proving its safe operation is challenging. While guaranteeing the absence of mistakes in a piece of software is not always possible, even the identification of their presence is not an easy task. Traditional quality assurance techniques like code reviews or test case generation are also insufficient in this case due to the critical nature of the medical domain. Additionally, the lack of domain knowledge of software engineers makes the matter worse. 

We present an approach where a system is synthesized using an incremental refinement process synchronizing and integrating different views and abstraction levels of the system. The process of quality assurance is embedded in the model development. Requirements of the system are supplied to a refinement-based development process that rigorously checks them for consistency and conformance. Every time a new requirement is specified, it first undergoes an internal consistency check and then, additionally, it is also confirmed with the stakeholders whether this requirement indeed captures the desired behavior. The stakeholders, in this way, become part of the development process right from the start and also the chance of an error to trickle down to the later stages of the development process is minimized. 

\begin{figure}
\centering
\includegraphics[width=0.9\linewidth]{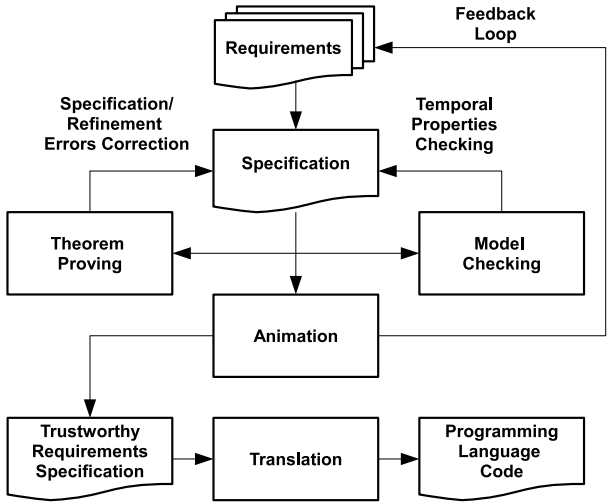}
\caption{The formal development paradigm}
\label{fig:approach}
\end{figure}

As shown in Figure \ref{fig:approach}, our approach for the development of high-assurance AMDs consists of four major steps:

\begin{enumerate}
\item formal requirements specification,
\item verification, 
\item validation, and
\item code generation.
\end{enumerate}

In the requirements specification step, informal user and system requirements are translated into a formal specification using a rigorous method. During this process, requirements are precisely written using mathematical and logical structures which are amenable to formal analysis to determine their correctness. 

One of the important cornerstones of the specification process is the representation of requirements at various abstraction levels using the notion of refinement. By following this technique, requirements are easy to specify, analyze and implement. In this style of specification writing, requirements are incrementally added to the model until the model is detailed enough to be effectively implemented.

Once the informal requirements have been translated into a formal specification, the next step is to make sure that the requirements conform to verification standards, i.e., requirements are consistent and verifiable. During this process, it is determined that a specification conforms to some precisely expressed properties that the model is intended to fulfill such as well-definedness, invariant preservation and other safety conditions. 

According to \cite{clarke96a}, two well-established formal verification approaches are theorem proving and model checking. While the former refers to reasoning about defined properties using a rigorous mathematical approach, the latter is the process of exploration of the whole state space of a model to verify dynamic properties. Both deductive theorem proving and model checking are important for proving the consistency of an AMD. While theorem proving is helpful in ensuring safety constraints of the system, model checking is effective in verifying temporal constraints of the system such as liveness and fairness properties.

Once a requirement is specified and verified, the next step to consider is its validation. Validation is a process where it is established by examination and provision of objective evidence that the stakeholders' requirements have been captured correctly and completely in the requirements specification document. Verification alone is not sufficient to guarantee correctness of the model because it does not check whether the specification documents the requirements from the viewpoints of stakeholders. 

In order to make stakeholders understand the formal specification, we propose to animate the specification. Animation is a process to demonstrate the fundamental operations of a specification using a dynamic and interactive graphical display. This technique is very well-suited for making a quick mental image of the model even for non-technical domain experts. 

The last step of the formal development process is the translation of the requirements specification into programmable code. The last refinement step of the specification writing process is, in fact, already very detailed and close to the implementation stage. An automatic code generation utility capable of translating a formal specification into the target language code such as \cite{wright09a} or \cite{singh13a} can be used at this step.

\section{Case Study: Hemodialysis Machines}
\label{sec:casestudy}

For the experimental validation of our approach, we apply our approach to a HD machine case study \cite{mashkoor15b}. The approach is applied by employing the formal method Event-B \cite{abrial10a} and its support platform Rodin \cite{abrial10b}. A typical specification in Event-B consists of two parts: states and events. A state is a mapping between names and values constrained by an invariant. An event is responsible for transitions between one state and another. For practical purposes, Event-B models are split into Contexts and Machines, each describing the constant and the variable part of the state respectively.


\subsection{Hemodialysis process}
\begin{figure}[htpb]
\centering
\includegraphics[width=0.9\linewidth]{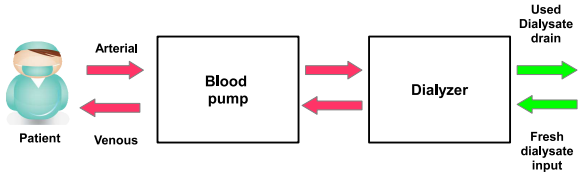}
\caption{Working principle of HD machines}
\label{fig:architecture}
\end{figure}

HD is a treatment for kidney failure that uses a machine to send the patient's blood through a filter, called a dialyzer, for extracorporeal removal of waste products. The blood is taken through the arterial access of the patient's body. The blood then travels through a tube that takes it to the dialyzer. Inside the dialyzer, the blood flows through thin fibers that filter out wastes and extra fluid using dialysate, a chemical substance that is used in HD to draw fluids and toxins out and to supply electrolytes and other chemicals to the bloodstream. The cleaned blood is then recycled back to the patient through the venous access. A vascular access lets large amounts of blood flow continuously during HD treatments to filter as much blood as possible per treatment. A specific amount of blood is conducted through the machine every minute. The working principle of HD machines is depicted by Figure \ref{fig:architecture}.

\subsection{Model development}

\subsubsection{Specification step}

The model has been developed using the following pattern:

\begin{enumerate}

\item Synchronize each requirement with a refinement step,
\item Distinguish and specify the static and dynamic elements of the requirement in the context and the machine of the related refinement respectively,
\item Introduce the safety properties expressed in the  requirement as machine invariants,
\item Introduce the monitoring events.
\end{enumerate}

The following is an example of how the requirements of the HD machine case study are specified in the Event-B model.

\begin{mdframed}
{\bf Requirement}: If the system is in the preparation mode and performs priming or rinsing or if the system is in the therapy mode and if the dialysate temperature exceeds the maximum temperature of 41$^{\circ}$C, then the software shall disconnect the dialyser from the dialysate within 60 seconds and execute an alarm signal.
\end{mdframed}

%

We first initiate a context that defines requirement-related static data such as modes, operations, and alarms. Then the corresponding machine of the refinement is specified. It contains several variables and invariants. The following invariants {\tt inv1} and {\tt inv2} specify the related safety requirements. 

\begin{lstlisting}
inv1 softwareMode = Preparation & (operation = Priming or operation = Rinsing)
      & dialysateTemperature > 41 => 
      dialyserState = {Dialysate |-> DialyserDisconnected} &
      dialyserDisconnectionTime < 60 & alarm = ALM377
inv2 softwareMode = Therapy & dialysateTemperature > 41 =>
      dialyserState = {Dialysate |-> DialyserDisconnected} &
      dialyserDisconnectionTime < 60 & alarm = ALM639
\end{lstlisting}

Then we specify two monitoring events to capture the behavior of the system because both events trigger different alarms. As shown in Figure \ref{fig:event}, the event {\tt disconnectDialyserPreparation} is triggered when the software is in the preparation mode and the temperature of the dialysate rises to more than 41$^{\circ}$C during the operation. However, if the software is in the therapy mode while the same thing happens, then the other monitoring event is triggered. 

\begin{figure}[htbp]
\begin{center}
  \begin{minipage}{0.8\linewidth}
\begin{lstlisting}[frame=single]
Event disconnectDialyserPreparation
  Where
   softwareMode = Preparation & dialysateTemperature > 41 &
   (operation = Priming or operation = Rinsing) &
   dialyserState = {Dialysate |-> DialyserConnected} &
   dialyserDisconnectionTime < 60
  Then
   dialyserState := {Dialysate |-> DialyserDisconnected} & 
   alarm := ALM377 & dialyserDisconnectionTime := 0
End
\end{lstlisting}
  \end{minipage}
\end{center}
\caption{Event {\tt disconnectDialyserPreparation}}
\label{fig:event}
\end{figure}

An additional event {\tt dialyserDisconnecitonClock} is also specified to monitor the timing constraint of the requirement. The tick of the clock is modeled as $\mathbb{N}$ that increments  {\tt dialyserDisconnectionTime} by 1 (second). 

\subsubsection{Verification step}

Verification of a model is achieved when it is proven that the model is free from specification errors and inconsistencies. Our fully-proven specification ensures that the model is consistent, well-defined and its events preserve its invariants. Additionally, we also prove that concrete events in later refinement steps maintain invariants of the abstract refinements, maintain abstraction invariants, and, when appropriate, decrease variants monotonically. 

Temporal properties (safety and liveness) of the system are checked by a combination of theorem proving, model checking and animation. Using theorem proving, we proved that safety invariants are preserved by the behavior of the system, the system is non-divergent, i.e., new events do not take control forever by preventing events from the abstract models from happening, and the system preserves enabledness, i.e., if an event is enabled
at an abstract level then it is also enabled at the concrete level. Using model checking, we ensured that legal states of the model are reachable, specified formulas are satisfiable and the model does not contain any deadlock. Using animation, we successfully checked that system traces eventually reach their intended final states.

\subsubsection{Validation step}

Validation of a model is achieved when it is demonstrated that the model is free from requirements errors and reflects the stakeholders' wishes adequately. The most common way to validate a specification in the Event-B method is to animate the specification by invoking its operational semantics to inspect its behavior. 

For animation and model checking of our specification, we have used the ProB tool \cite{leuschel08a} that
supports automated consistency checking of Event-B machines via  constraint solving techniques. Animation using ProB worked very well. We created several behavioral scenarios and executed them accordingly to demonstrate the behavior of the system to stakeholders. The ProB tool assisted us in finding potential invariant problems and their improvement by generating counterexamples whenever an invariant violation is discovered. ProB also helped us in improving invariant expressions by providing hints for strengthening invariants each time an invariant was modified or a new proof was generated by the Rodin platform. We corrected more errors during specification modeling and reviewing than during discharging proofs and animation. 

\subsubsection{Code generation step}
At the last refinement step, when the specification is sufficiently detailed, we extract the programing language code out of it. This is the result of the translation process that converts the B code into a sequential programing language code that runs on the given hardware. Our target language is C. We use the EB2All tool \cite{singh13a} for the translation purpose. 

\section{Discussion, conclusion and future work}
\label{sec:conc}

This letter focuses on the rigorous development of a software-controlled safety-critical AMD responsible for renal replacement therapy. Our employed approach successfully enabled us to specify and analyze various critical components of HD machines at different abstraction levels. Our approach also enabled us to detect and correct errors and omissions close to the point of their introduction. The formal Event-B method supported by the Eclipse-based open source Rodin tool has also lent itself to the development of such systems. We have found Event-B an adequate method for the modeling and analysis of critical medical devices. Its refinement principles and V\&V mechanisms provide all the elements that are necessary for the safe development of AMDs and are in full accordance with IEC standards and FDA guidelines. 

However, during the development, we also faced several challenges. For example, sophisticated tools and elaborated guidelines for managing the complexity of growing models by decomposition are missing in Event-B and Rodin. There is also no implicit notion of time in Event-B, although this will be necessary for an elegant expression of timing properties which play a critical role in medical devices. Currently, we resort to ProB for proving various temporal properties of the system, but ProB often fails (at the detailed level of refinements) due to the state space enumeration and explosion problems. In our opinion, a standard and more natural way is required to specify and prove that temporal properties of the system are preserved by Event-B refinements. Finally, a tool that is able to generate {\it ready-to-deploy} machine code from Event-B formal models is also missing; currently available tools can not do this. The notable problems with these tools is that the Event-B model must be restricted to a well-defined subset in order to generate C code and a formal proof that the translation process preserves the safety properties of the model is missing. Extensions of available code generation tools in these directions is an issue for future work.  

\bibliographystyle{IEEETran}

\end{document}